\documentclass{iopart}
\usepackage{amsmath,amssymb,bbm,graphicx}
\begin{document}
\title{Generalized measurement of the non-normal two-boson operator $Z_\gamma
= a_1 + \gamma a_2^\dag$}
\author{Matteo G A Paris\footnote{\tt matteo.paris@fisica.unimi.it}$^{1,2}$, 
Giulio Landolfi\footnote{\tt giulio.landolfi@le.infn.it}$^{3,4}$, 
Giulio Soliani\footnote{\tt giulio.soliani@le.infn.it}$^{3,4}$}
\address{$^1$Dipartimento di Fisica dell'Universit\'a di Milano, I-20133 Milano, Italy \\ 
$^2$Institute for Scientific Interchange Foundation, I-10133 Torino, Italy\\
$^3$Dipartimento di Fisica dell'Universit\'a del Salento, I-73100 Lecce, Italy \\ 
$^4$Istituto Nazionale di Fisica Nucleare, Sezione di Lecce, I-73100 Lecce, Italy}
\pacs{03.65.Ta, 42.50.Xa}
\begin{abstract}
We address the generalized measurement of the two-boson operator
$Z_\gamma= a_1 + \gamma a_2^\dag$ which, for $|\gamma|^2 \neq 1$, is not
normal and cannot be detected by a joint measurement of quadratures on
the two bosons. We explicitly construct the minimal Naimark extension,
which involves a single additional bosonic system, and present its
decomposition in terms of two-boson linear SU(2) interactions. The
statistics of the measurement and the added noise are analyzed in
details. Results are exploited to revisit the Caves-Shapiro concept of
generalized phase observable based on heterodyne detection. 
\end{abstract}
The two-boson operator
\begin{align}
Z_\gamma=a_1 + \gamma a_2^\dag
\label{Z}\:,
\end{align}
is normal $\left[Z_\gamma,Z_\gamma^\dag\right]= 1-|\gamma |^2$  
for $|\gamma|=1$. In this case the real $X_\gamma = \frac12 (Z_\gamma +
Z_\gamma^\dag)$ and the imaginary $Y_\gamma= \frac{i}2 (Z_\gamma -
Z_\gamma^\dag)$ parts of $Z_\gamma$ commute $\left[X_\gamma,Y_\gamma\right]=0$ 
and can be jointly measured. Actually they correspond to the canonical sum- and 
difference-quadratures of the two modes
\begin{align}
X_\gamma = \frac1{\sqrt{2}} \left( q_1 + q_2 \right) \quad 
Y_\gamma = \frac1{\sqrt{2}} \left( p_1 - p_2 \right)  
\label{ZRZI}\:,
\end{align}
where, for $k=1,2$,  
\begin{align}
q_k = \frac{1}{\sqrt{2}} \left(a_k^\dag + a_k\right) \quad 
p_k = \frac{i}{\sqrt{2}} \left(a_k^\dag - a_k\right) \quad [q_j,p_k]=i
\delta_{jk} 
\label{defQ}\:.
\end{align}
On the other hand, for $|\gamma| \neq 1$, we have
\begin{align}
X_\gamma  = \frac1{\sqrt{2}} \left( q_1 + |\gamma | x_{2,\theta_\gamma} \right) \quad
Y_\gamma  = \frac1{\sqrt{2}} \left( p_1 - |\gamma | x_{2,\theta_{\gamma}+\pi/2}\right)
\label{PQgnot1}
\end{align}
where $x_{k,\phi} = \frac1{\sqrt{2}} (a_k^\dag e^{i\phi} + a_k e^{-i\phi})$ is a
rotated quadrature of the $k$-th boson and $\theta_\gamma = \arg \gamma$.
In this case, the two operators do no commute
$\left[X_\gamma,Y_\gamma\right] = \frac{i}{2} (1-|\gamma |^2)$ 
and a generalized measurement should be devised.
Indeed, the eigenstates of $Z_\gamma$ for $\gamma\neq 1$  
$$|z\rangle\rangle_\gamma = D(z)\otimes {\mathbb I}\: 
|\gamma\rangle\rangle$$ 
where $D(z) = \exp\{z a_1^\dag - z^* a_1\}$ is the displacement operator 
and $|\gamma\rangle\rangle =  \sqrt{1-|\gamma |^2} \sum_n \gamma^n 
|n\rangle\otimes |n\rangle$, {\em do not} provide a resolution of the
identity, we have 
$$\int \frac{d^2 z_\gamma}{\pi}\: |z\rangle\rangle_\gamma {}_\gamma\langle\langle z | =
(1-|\gamma |^2) |\gamma |^{2 a^\dag a}\:.$$
\par
We first notice that $Z_\gamma = R_{\theta_\gamma}^\dag Z_{|\gamma|}
R_{\theta_\gamma}$ where $R_\phi = \exp (i\,\phi\,a_2^\dag a_2)$ and therefore,
without loss of generality, we may restrict attention to the case of real
positive $\gamma$. In this case we have 
\begin{align}
X_\gamma  = \frac1{\sqrt{2}} \left( q_1 + \gamma  q_2  \right) \quad
Y_\gamma  = \frac1{\sqrt{2}} \left( p_1 - \gamma  p_2 \right)
\label{PQrealg}
\end{align}
In addition, we notice that, up to a permutation of the mode labels,  
$Z_\gamma = \gamma Z^\dag_{\gamma^{-1}}$ and therefore, since the 
multiplicative constant does not influence the measurement scheme, we 
may further restrict attention to the case $0 < \gamma < 1$.
\par
The operator $Z_\gamma$ is defined on the Hilbert-Fock space ${\cal H}_{12}$ 
of two harmonic oscillators. A Naimark extension for the operator $Z_\gamma$ 
is a triplet $\left({{ \cal H}_a, T_\gamma, \sigma}\right)$, 
where $T_\gamma$ is an operator  defined on an extended Hilbert space ${\cal
H}_{12}\otimes {\cal H}_a$ and $\sigma$ is a state (density operator) in 
${\cal H}_a$, such that for any state $R\in {\cal H}_{12}$ we have
\begin{align}
\hbox{Tr}_{12}\left[R\: X_\gamma \right] & =
\hbox{Tr}_{12a}\left[R\otimes\sigma\: \hbox{Re }T_\gamma \right] 
\nonumber \\
\hbox{Tr}_{12}\left[R\: Y_\gamma \right] & =
\hbox{Tr}_{12a}\left[R\otimes\sigma\: \hbox{Im }T_\gamma \right]
\label{Naimark}\:.
\end{align}
Equations (\ref{Naimark}) are usually summarized by saying that the operator
$T_\gamma$ {\em traces} the operator $Z_\gamma$.
Of course, Eqs. (\ref{Naimark}) do not hold for higher moments: the 
generalized measurement of $Z_\gamma$ unavoidably introduces some noise 
of purely quantum origin. In general we have 
\begin{align}
\hbox{Tr}_{12}\left[R\: X_\gamma^n \right]  & \neq 
\hbox{Tr}_{12a}\left[R\otimes\sigma\: (\hbox{Re }T_\gamma)^n \right]
\quad n\geq 2
\nonumber \\ 
\hbox{Tr}_{12}\left[R\: Y_\gamma^n \right]  & \neq 
\hbox{Tr}_{12a}\left[R\otimes\sigma\: (\hbox{Im }T_\gamma)^n \right]
\quad n\geq 2\:.
\label{nl2}
\end{align}
In this communication we look for a minimal Naimark extension, that is 
an extension involving a single additional bosonic mode $a_3$.  In general, 
for operator of the form $T_\gamma = Z_\gamma + f(a_3,a_3^\dag)$ the trace 
condition of Eqs. (\ref{Naimark}) require $\hbox{Tr}_a \left[\sigma\,
f(a_3,a_3^\dag)\right]=0$, whereas the constraint of normality can be 
written as 
\begin{align}
0\equiv \left[T_\gamma,T_\gamma^\dag\right] =
\left[Z_\gamma,Z_\gamma^\dag\right] + \left[f(a_3,a_3^\dag),f(a_3,a_3^\dag)^\dag\right]
\label{normalT}\:.
\end{align}
It is straightforwardly seen that 
$f(a_3,a_3^\dag)=\kappa a_3$ or $f(a_3,a_3^\dag)=\kappa a_3^\dag$, 
where $\kappa$ is a real constant, are solutions of 
Eqs. (\ref{Naimark}) and (\ref{normalT}). 
In the following we analyze in details whether this kind of extensions can be
implemented {\em using only bilinear interactions} 
among the three modes followed by measurement of quadratures at the output. 
\par
The measurement scheme is the following: the modes $a_k$ interact each other 
via the unitary operator $U_\gamma$,  which impose the linear transformation
\begin{align}
\label{evolution}
\left(\begin{array}{c} A_1 \\ A_2 \\ A_3 \end{array}\right)&
=U_\gamma^\dag 
\left(\begin{array}{c} a_1 \\ a_2 \\ a_3 \end{array}\right)
U_\gamma = {\mathbf M}
\left(\begin{array}{c} a_1 \\ a_2 \\ a_3 \end{array}\right)
\end{align}
and then, at the output, the quadratures 
$$Q_1= \frac1{\sqrt{2}} (A_1 + A_1^\dag) \quad P_2=\frac{i}{\sqrt{2}}(A_2^\dag-A_2)$$ 
are measured with the aim of obtaining, upon the definition 
$T_\gamma = Q_1 + i P_2$,
\begin{align}
\hbox{Tr}_{12}\left[R\: X_\gamma \right] & =
\hbox{Tr}_{12a}\left[R\otimes\sigma\: Q_1\right] \\
\hbox{Tr}_{12}\left[R\: Y_\gamma \right] & =
\hbox{Tr}_{12a}\left[R\otimes\sigma\: P_2\right]
\label{Naimark2}\:
\end{align}
for {\em any} $R$, and at least one $\sigma$ such that $\hbox{Tr}[\sigma\,a_3]=0$. 
A suitable evolution operator $U_\gamma$ corresponds to the transformation 
\begin{align}
{\mathbf M} = \frac{1}{\sqrt{2}}\left( 
\begin{array}{ccc}
1 &   \gamma & \kappa \\
1 &  - \gamma & - \kappa \\
m_1 & m_2 & m_3 
\end{array}
\right)\:.
\label{matrixM}
\end{align}
Upon imposing the constraint of unitarity, {\em i.e}
$[A_j,A_k^\dag]=\delta_{jk}$, we have the solution
\begin{align}
\kappa & = \sqrt{1 -\gamma^2} \, , \:\: \quad \qquad m_1 =0 \, ,\\
m_2  & = -\sqrt{2 (1-\gamma^2)} \, , \quad \quad  m_3 = \sqrt{2} \gamma \, ,
\label{kappa}
\end{align}
which makes ${\mathbf M}$ a $U(3)$ transformation and leads to 
\begin{align}
Q_1 = \frac{1}{\sqrt{2}} \left( q_1 + \gamma q_2 + \sqrt{1-\gamma^2} q_3
\right) \: , \quad
P_2 = \frac{1}{\sqrt{2}} \left( p_1 - \gamma p_2 - \sqrt{1-\gamma^2} p_3 \right)
\label{quadQP}\:,
\end{align}
and, in turn, to $T_\gamma = a_1 + \gamma a_2^\dag + \kappa a_3^\dag$.
Notice that no unitary solution can be found (for $|\gamma | <1$)
for the case
$f(a_3,a_3^\dag)=\kappa a_3$, {\em i.e.} for linear transformation
expressing the output modes $(A_1,A_2,A_3)$ as a linear combination
of $(a_1,a_2,a_3^\dag)$ \footnote{Actually, a solution involving a
SU(1,1) interaction between $a_2$ and $a_3$ followed by a SU(2)
interaction between $a_1$ and $a_2$ may be found for $|\gamma |>1$
and then extended to the whole range of $|\gamma |$ by rescaling.
However, this solution unavoidably introduces a larger amount of
noise compared to that of Eqs. (\ref{ChiG}) and (\ref{vacvac}) and
it will not be considered here.}
\par
\begin{figure}[ht!]
\centerline{\includegraphics[width=0.85\textwidth]{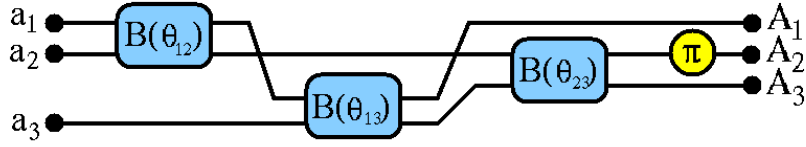}}
\caption{Block diagram of the decomposition of the ${\mathbf M}$
transformation of Eq. (\ref{matrixM}) into three SU(2) transformations, each
involving two of the modes, plus a $\pi$-rotation. The boxes corresponds
to evolution operators of the form $B_{jk}(\theta_{jk}) =
e^{- i \theta_{jk} \left(a_j a_k^\dag + a_k a_j^\dag\right)}$
(see text).}
\label{f:decM}
\end{figure}
\par
A question arises on how the unitary $U_\gamma$ can be implemented in practice,
as for example in a quantum optical setting. As it is well known, any SU(3)
transformation may be decomposed into a set of SU(2) transformation \cite{dec}.
In our case the U(3) ${\mathbf M}$-transformation may be decomposed using
three SU(2) transformations followed by a $\pi$-rotation. In Fig. \ref{f:decM}
we report the explicit decomposition of ${\mathbf M}$. The circle denotes a
$\pi$-rotation on the second mode {\em i.e.} a unitary of the form
$R_2=\exp\{i\pi a_2^\dag a_2\}$. The boxes correspond to SU(2) rotations
{\em i.e.} to evolution operators of the form $B_{jk}(\theta_{jk}) =
\exp\left\{ - i \theta_{jk} \left(a_j a_k^\dag + a_k a_j^\dag\right)\right\}$,
corresponding to the transformations
\begin{align}
B_{jk}^\dag(\theta_{jk})
\left(\begin{array}{c} a_j \\ a_k \end{array}\right)&
B_{jk}(\theta_{jk})=
\left(
\begin{array}{cc}
\cos\theta_{ij} & \sin\theta_{ij} \\
-\sin\theta_{ij}& \cos\theta_{ij}
\end{array}
\right)
\left(\begin{array}{c} a_j \\ a_k \end{array}\right)&
\label{bs}
\end{align}
By explicit construction we have
$$ U_\gamma =
\left[{\mathbbm I}_1 \otimes R_2 \otimes {\mathbbm I}_3 \right]
\left[B_{23}(\theta_{23}) \otimes {\mathbbm I}_3 \right]
\left[B_{13}(\theta_{13}) \otimes {\mathbbm I}_2 \right]
\left[B_{12}(\theta_{12}) \otimes {\mathbbm I}_1 \right]
$$
where
\begin{align}
\cos\theta_{23}=\sqrt{\frac{1+\gamma^2}{2}}\: ,\quad
\cos\theta_{13}=\sqrt{\frac{2\gamma^2}{1+\gamma^2}} \: , \quad
\cos\theta_{12}=\sqrt{\frac{\gamma^2}{1+\gamma^2}}\: .
\end{align}
Other decompositions may be also found, allowing for permutations of modes
and different rotations.
For $\gamma \rightarrow 1$ the mode $a_3$ decouples from the other
two modes and the scheme reduces to the joint measurement of quadratures
for the normal operator $Z_1$ \cite{wal}.
\par
Each outcome from the joint measurement of the quadratures
$Q_1$ and $P_2$ corresponds to a complex number $\tau=Q_1 + i P_2$
that represents a realization of the observable $T_\gamma$.
The probability density
of the outcomes $K_\gamma(\tau)$ for a given initial preparation
$R\otimes\sigma$ is obtained as the Fourier transform of the
moment generating function $\Xi(\lambda )$
\begin{align}
K_\gamma(\tau) = \int \frac{d^2\lambda}{\pi^2} e^{\lambda^* \tau -
\lambda\tau^*}\: \Xi (\lambda)\:,
\end{align}
where
\begin{align}
\Xi (\lambda) =
\hbox{Tr}\left[R\otimes\sigma\: e^{\lambda T^\dag_\gamma -
\lambda^*  T_\gamma}\right]\:.
\label{Chi}
\end{align}
Using Eqs. (\ref{quadQP})  we have
$\exp\{\lambda T^\dag_\gamma -
\lambda^*  T_\gamma\} =
 D_1(\lambda)\otimes D_2(-\lambda\gamma)\otimes D_3 (-\lambda \kappa)$
 where $D_j(z)$ is the displacement operator
for the mode $a_j$. Therefore, the moment generating function rewrites as
\begin{align}
\Xi_\gamma (\lambda) =
\chi_{12} (\lambda)\: \chi_3 (-\lambda \kappa)
\label{Chipm}\:,
\end{align}
where
$\chi_{12} (\lambda)=\hbox{tr}\left[R\:D_1(\lambda)\otimes
D_2(-\lambda\gamma)\right]$ and
$\chi_3 (z) = \hbox{Tr}[\sigma\: D_3(z)]$ is the characteristic function
of the mode $a_3$.
Using (\ref{Chipm}) is easy to see that the probability density of
the outcomes is given by  the convolution
\begin{align}
K_\gamma(\tau)= \frac{1}{\kappa^2}
H_\gamma (\tau) \star  W_3 (-\tau/\kappa)
\label{Kgm}\:,
\end{align}
$W_3(z)$ being the Wigner function of the mode $a_3$, 
$\star$ the convolution product, and
$H_\gamma (z)$ the density obtained by the
Fourier transform of $\chi_{12} (\lambda)$.
In turn, for factorized preparations $R=\varrho_1\otimes\varrho_2$
the moment generating function $\chi_{12}(\lambda)= \chi_1 (\lambda)\:
\chi_2 (-\lambda)$ factorizes into the product of the
characteristic functions of $\varrho_1$ and $\varrho_2$ respectively,
and the density $H_\gamma(\tau)$ reduces to the convolution of
the Wigner functions of the two input signals
\begin{align}
H_\gamma(\tau) =\frac1{\gamma^2}\:W_1 (\tau) \star W_2 (-\tau/\gamma)\:.
\end{align}
Using (\ref{quadQP}) it is straightforward to 
see how the variances of the measured quantities $Q_1$ and $P_2$ are 
related to the variances of the quadratures of interest. We have 
\begin{align}
\Delta Q_1^2 &= \Delta X_\gamma^2 + 
\frac12 (1-\gamma^2) \Delta q_3^2   \nonumber \\
\Delta P_2^2 &= \Delta Y_\gamma^2 + 
\frac12 (1-\gamma^2)\Delta p_3^2 \:, \label{addN}
\end{align}
where $\Delta q_3^2 = \hbox{Tr}[\sigma\:q_3^2]$
and analogously $\Delta p_3^2 = \hbox{Tr}[\sigma\:p_3^2]$ 
(remind that Eq. (\ref{Naimark}) implies
$\hbox{Tr}[\sigma\:q_3]=\hbox{Tr}[\sigma\:p_3]=0$). 
Notice that the added noise in Eq. (\ref{addN}) is the minimum 
noise according to generalized uncertainty 
relations for joint measurement of non commuting observables
\cite{art1,GL,art2,yue,bus}.
On the other hand, the covariance 
between the measured quadratures {\em i.e.} the
quantity
\begin{align}
\Sigma_{Q_1 P_2} = \frac12 \hbox{Tr} _{12a}\left[ 
R\otimes\sigma\: (Q_1 P_2 + P_2 Q_1)
\right] - 
\hbox{Tr}_{12a}\left[R\otimes\sigma\:Q_1 \right]
\hbox{Tr}_{12a}\left[R\otimes\sigma\:P_2 \right]
\:,\label{cov1}
\end{align}
may be written as
\begin{align}
\Sigma_{Q_1 P_2} =  \Sigma_{X_\gamma Y_\gamma} - 
\frac12 (1-\gamma^2) \hbox{Tr}_a\left[\frac12 
\sigma \left(p_3 q_3 + q_3 p_3 \right)\right]
\label{cov2}
\end{align}
where
$\Sigma_{X_\gamma Y_\gamma}= \frac12 \hbox{Tr} _{12}\left[ 
R\: (X_\gamma Y_\gamma + Y_\gamma X_\gamma) \right] - \hbox{Tr}_{12}
\left[R\:X_\gamma \right]\hbox{Tr}_{12}\left[R\:Y_\gamma \right]$ 
is the covariance of the desired quadratures.
\par
Notice that the added noise to the covariance, Eq. (\ref{cov2}), may
vanish for some preparation of the state $\sigma$ whereas the added
noise to the variances, Eq. (\ref{addN}), cannot vanish for any physical
preparation $\sigma$. This raises the question of the consequences of
different field states on the statistics of the measurement and, in
turn, of the role played  by preparations of states in  concrete
experiments.  On the other hand, within experimental frameworks, one may
take full advantage of possible freedom in preparing some of the modes.
This is definitively the case of the Naimark mode $a_3$, even though its
preparation needs to be compatible with the prescription
(\ref{Naimark}) for  the expectation values of position and momentum
operators.  In particular, a valid Naimark extension can be obtained by
preparing the mode $a_3$ in the vacuum state $\sigma=|0\rangle\langle
0|$ to let its contribution to the noise in formula (\ref{cov2}) to
vanish, since $\hbox{Tr}_a [ \sigma (q_3 p_3+p_3q_3) ]=0$, and to minimize
$\Delta q_3^2$ and $\Delta p_3^2$ in (\ref{addN}), since both the terms
would be equal to one half.  Each of the other two fields may be, for
instance, in one among the most meaningful types of states, such as
number states, coherent states, thermal states or phase states (i.e.
eigenstates of the operator $C+iS$, where $C$ and $S$ are ``cosine'' and
``sine'' operators respectively) or prepared in an entangled states. 
If we consider the fully separable state described by the density operator
$\varrho=R \otimes \sigma =\varrho_1 \otimes \varrho_2 \otimes \sigma$, 
where $\varrho_k$, with $k=1,2$, denotes the preparation for the k-th
bosonic field in the arbitrarily mixed state 
$\varrho_k=\sum^{\infty}_{m=0} p_m^{(k)} |m \rangle \langle m |$ 
on the Hilbert space ${\cal H}_k$, then the system moment generating 
function is easily obtained by resorting to 
\begin{align}
\hbox{Tr}_k \left[ \varrho_k D_k (\alpha_k)\right]= 
e^{-\frac{|\alpha_k|^2}{2}}
\sum_{m=0}^{\infty} p_m^{(k)} L_{m} \left( | \alpha_k |^2 \right) \quad, 
\label{Trpmk} 
\end{align}
where the $L_n$'s are Laguerre polynomials. 
For instance, for coherent and phase states Eq. (\ref{Trpmk}) 
should be used with 
\begin{align}
p_m^{(k)}=e^{-|\alpha|^2} \frac{|\alpha|^{2m}}{m!}
\qquad \hbox{and} \qquad
p_m^{(k)}=(1-|z|^2) |z|^{2m} 
\end{align}
respectively (phase state formulae can be used even when dealing with
thermal states upon the identification $z=\exp[-\frac{1}{2}\beta \hbar
\omega]$, $\beta$ being the inverse of temperature).  Suppose no
specific conditions do constraint, in principle, the preparation for the
mode $a_2$. Once again a vacuum choice may be advantageous in some
respects.   Let us therefore focus on the specific case of the
measurement of $Z_\gamma$ on the class of factorized signals described
by $R=\varrho_1 \otimes |0\rangle\langle 0|$ where $\varrho_1$ is a
generic preparation of the mode $a_1$ while $|0\rangle$ is the ground
state of the mode $a_2$.  In this case $\varrho=\varrho_1 \otimes
|0\rangle\langle 0 | \otimes  |0\rangle\langle 0|$, Eq. (\ref{Kgm})
becomes a Gaussian convolution and the moment generating function
becomes independent of the parameter $\gamma$
\begin{align}
\Xi (\lambda) = 
\chi_1 (\lambda)\: \exp \left(-\frac12 |\lambda|^2 \right)
\label{ChiG}\:.
\end{align}
The measured variances are thus given by 
\begin{align}
\Delta Q_1^2  =  \frac12 \left( \Delta q_1^2 + 1 \right)\qquad 
\Delta P_2^2  =  \frac12 \left( \Delta p_1^2 + 1 \right) 
\label{vacvac}
\end{align}
Equations (\ref{ChiG}) and (\ref{vacvac})
contain a {\em remarkable result} that may be expressed as follows.
The measurement of $Z_\gamma$ on the class
of states $R=\varrho_1 \otimes |0\rangle\langle 0|$ {\em does not}
lead to added noise with respect to the measurement of the normal
operator $Z_1$. 
\par
\begin{figure}[ht!]
\centerline{\includegraphics[width=0.7\textwidth,height=0.14\textheight]{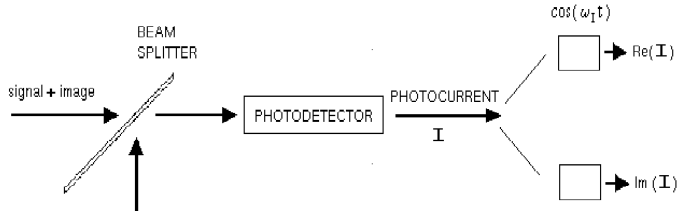}}
\caption{The scheme for heterodyne detection.}
\label{f:hdet}
\end{figure}
This result finds a natural application in the context of {\em
heterodyne detection},  where currents of the form (\ref{Z}) show up.
As it is known, in heterodyne detection a single-mode signal field
$E_{1}$ of nominal frequency $\omega _{1}$ is mixed through a
beam-splitter with a local oscillator field $E_{L}$ whose frequency
$\omega _{L}$ is slightly offset by an amount $\omega _{I} \ll \omega_1$
from that of the input signal, i.e. $\omega _{1}=\omega _{L}+\omega
_{I}$.  A photodetector is placed right after the beam-splitter (see
Fig. \ref{f:hdet}).  The output photocurrent, which generally depends on
fields parameters  and on specific assumptions on the apparatus, is
filtered at the \textit{intermediate frequency} $\omega _{I}$.  In
standard optical heterodyne detection (see e.g. \cite{shap}), measuring
the filtered photocurrent corresponds to realize the quantum measurement
of the normal operator $y = a_{1}+a_{2}^{\dag }$ \cite{shap}, where
$a_1$ (res. $a_2^{\dag})$ denotes the photon annihilator (resp.
creation) operator for the input (resp. image) signal.  Measuring the
real and imaginary parts of the (actually rescaled) output photocurrent
thus provides the simultaneous measurement of both input field
quadratures.  Nevertheless, it has been also argued that whenever one is
not restricted to an input field frequency in the optical regime, but,
rather, one is concerned with microwave (or radio) heterodyning, then
the interaction of the input signal field with the apparatus of Fig. 2.
(approximatively) results in the measurement operator 
$y_C= \sqrt{(1+\frac{\omega_I}{\omega_1})} a_1+\sqrt{(1-
\frac{\omega_I}{\omega_1})}  a_2^{\dag}$
(see \cite{cav} and discussion in \cite{shap}).  Since
$[y_C,y_C^{\dag}]= 2\frac{\omega_I}{\omega_1}\neq 0 $, Caves measurement
operator $y_C$ is not compatible with simultaneous measurements of
signal quadratures. In other words, standard heterodyne detection 
cannot achieve the measurement of the Caves operator and a question
arises on whether simultaneous phase and amplitude measurements may be
accomplished in this case. The answer may be found in the results
reported above. In fact, the measurement of the Caves operator
corresponds to the generalized measurement of the non-normal operator
\begin{equation}
Z_{\gamma _{C}}=a_{1}+\gamma _{C}\,\,a_{2}^{\dag }\quad ,\quad \gamma _{C}=%
\sqrt{\frac{\omega_1 -\omega_{I}}{\omega _{1}+\omega _{I}}}\,\,<1\quad  
\label{Zcaves}
\end{equation}
In the light of our previous results, we thus learn that the
simultaneous measurement of the field quadratures for a
quasi-monochromatic signal can be realized even in the case when the
heterodyne apparatus yields a measurement operator of the Caves type,
Eq. (\ref{Zcaves}). To this aim, it suffices to generalize the
heterodyne detection scheme by introducing a single boson Naimark mode
and letting it interact with the other modes through the linear
transformation (\ref{evolution}).  Moreover, a suitable preparation
enables one to avoid additional noise with respect to that resulting in
the measurement of signal field quadratures within the framework of the
standard optical heterodyne detection.  
\par
It is worth also discussing the matter from the point of view of phase
operators since our results can be used  to proceed in defining a {\em
feasible phase} within the Caves description of heterodyning.  Since the
operator $T$ is normal, then its associated self-adjoint phase operator
\begin{align}
\theta _{T}=\frac{1}{2i}\ln \frac{T}{T^{\dag }}
\label{thetaT}
\end{align}
can be defined unambiguously indeed so that cosine and sine
quadrature operators
\[
C=\frac{1}{2}\left( \e^{i \theta_T}+e^{-i \theta_T}\right)\quad, \qquad
S=\frac{1}{2i}\left( e^{i \theta_T}-e^{-i \theta_T}\right) \quad
\]
obey the correct relation $C^2+S^2=1$.
It is now in order to recalling that the two-modes relative number
state representation discussed
by Ban (see \cite{ban} and Refs. therein) fits fairly with the
feasible phase concept of Shapiro and Wagner
(namely, the shift phase operator associated with the
Shapiro-Wagner measurement operator $y=a_1+a_2^{\dag}$).
Upon 
defining the 3-mode relative number operator $N=N_1-N_2-N_3$, where
$N_{k}=a_{k}^{\dag }a_{k}$ ($k=1,2,3$), one gets
\begin{align}
\left[ e^{i\theta _{T}},N\right] =e^{i\theta _{T}}\,, 
\qquad [N,\theta_T]=i \,.
\label{commDTN}
\end{align}
These relations are what one expects for genuine phase operators.
In other words,  a feasible phase can be naturally defined even in
the Caves description of heterodyning at the cost of  introducing of
a Naimark mode and generalizing  the 2-modes relative
state representation to a 3-modes one.
The commutator $[N,\theta_T]$ can then be
interpreted as the canonical conjugation
of the feasible phase for Caves heterodyne measurement operator
with respect to the operator mode number difference $N$.
\par
As final comments, notice that tracing out the Naimark mode $a_{3}$, and 
introducing symmetric ordering when needed in Eqs. (\ref{thetaT})-(\ref{commDTN}),
formulae given in \cite{lan} are recovered.
Further, it would be of interest to move towards the direction of
generalizing the relative number state representation for the description
of the phase operator of the generalized heterodyne measurement we have
introduced in this communication, and more generally for operators describing
linear amplifiers involving more than three modes.
This is also concerned with the investigation of the possibility to extract
basic algebraic structures underlying these systems to generalize algebras
given in \cite{lan}. These issues are currently under investigation and
results will be reported elsewhere.
\par\noindent\\
This work has been supported by MIUR through the projects 
PRIN-2005024254-002 and PRIN-SINTESI. 


\section*{References}
\begin{thebibliography}{99}
\bibitem{art1} E. Arthurs, J. L. Kelly, Bell. Syst. Tech. J. {\bf 44}, 725
(1965)
\bibitem{GL} J. P. Gordon, W. H. Louisell in {\em Physics of Quantum 
Electronics} (Mc-Graw-Hill, NY, 1966).
\bibitem{art2} E. Arthurs, M. S. Goodman, Phys. Rev. Lett.
{\bf 60}, 2447 (1988)
\bibitem{yue} H. P. Yuen, Phys. Lett. {\bf 91A}, 101 (1982)
\bibitem{bus} P. Busch, D. B. Pearson, preprint ArXiv:math-ph/0612074. 
\bibitem{dec} M. Reck et al., Phys. Rev. Lett. {\bf 73}, 58 (1994).
\bibitem{wal} N. G. Walker, J. E. Carrol, Opt. Quantum Electr.
{\bf 18}, 355 (1986); N. G. Walker, J. Mod. Opt. {\bf 34}, 16 (1987).
\bibitem{shap} J. H. Shapiro and S. S. Wagner, 
IEEE J. Quant. Electron. {\bf 20}, 803 (1984).
\bibitem{cav}C.M. Caves, Phys. Rev. {\bf D26}, 1817 (1982).
\bibitem{lan}G. Landolfi, G. Ruggeri and G. Soliani, Int. J. Mod.
Phys. {\bf B19}, 2287 (2005).
\bibitem{ban}  M. Ban,  Phys. Rev. {\bf A50}, 2785 (1994).
\end{thebibliography}
\end{document}